
\input phyzzx
\hfuzz 20pt
\font\mybb=msbm10 at 10pt

\def\Bbb#1{\hbox{\mybb#1}}

\def\bR{\Bbb {R}}

\def\bfomega{\omega\kern-7.0pt \omega}

\magnification=900



\def\C{\mkern1mu\raise2.2pt\hbox{$\scriptscriptstyle|$}\mkern-7mu{\rm C}}

\def\pd{\partial_}
\def\pu{\partial^}

\def\m{\mu}
\def\n{\nu}
\def\l{\lambda}

\def\p{\rho}
\def\s{\sigma}
\def\t{\tau}

\def\f{\phi}
\def\d{\delta}
\def\e{\epsilon}


\font\mybb=msbm10 at 12pt

\def\Bbb#1{\hbox{\mybb#1}}

\def\bR{\Bbb {R}}

\def\bfomega{\omega\kern-7.0pt \omega}


\REF\strom{A. Strominger and C. Vafa, {\sl Macroscopic Origin of the
Bekenstein-Hawking Entropy}, Phys. Lett. {\bf B379} 
(1996) 99, hep-th/9601029.}
\REF\ferrara{A.C. Cadavid, A. Ceresole, R. D'Auria and S. Ferrara,{\sl 
Eleven-Dimensional Supergravity Compactified of Calabi-Yau Threefolds },
Phys. Lett. {\bf B357} (1995) 76.}
\REF\gpapas{G. Papadopoulos and P.K. Townsend, {\sl Compactifications of D=11
Supergravity on Spaces of Exceptional Holonomy},  Phys. Lett.{\bf B357}
(1995) 300, hep-th/9506150.}
\REF\grwest{M.B. Green, J.H. Schwarz and P.C. West, {\sl Anomaly Free
Chiral Theories in Six Dimensions}, Nucl. Phys.
{\bf B 254} (1985) 327.}
\REF\town{M. G\"unaydin, G. Sierra and P.K. Townsend,  {\sl Exceptional
Supergravity Theories and the Magic Square}, Phys. Lett. {\bf B133}
(1983) 72; {\sl The Geometry of N=2 Maxwell-Einstein
Supergravity and Jordan Algebras}, Nucl. Phys. {\bf B242} (1984) 244.}
\REF\wit{B. de Wit and A. van Proeyen, {\sl Broken 
Sigma Model Isometries in
Very Special Geometry}, Phys. Lett. {\bf 293B} (1992) 95, hep-th/9207091.}
\REF\sabra{W.A. Sabra, {\sl General BPS Black Holes
in Five Dimensions}, Mod. Phys. Lett. {\bf A 13} 
(1998) 239, hep-th/9708103; 
A. Chamseddine and W.A. Sabra, {\sl Metrics admitting Killing 
spinors in five dimensions}, Phys. Lett. {\bf B426} (1998) 36.}
\REF\kalloshc{A. Chou, R. Kallosh, J. Rahmfeld, S-J. Ray, M. Shmakova
and W. K. Wong, {\sl Critical Points and Phase transitions
in 5-D compactifications of M-theory}, Nucl. Phys.
 {\bf B508} (1997) 147; hep-th/9704142.}
\REF\cfgk{A.H. Chamseddine, S. Ferrara, G.W. Gibbons and R. Kallosh,
{\sl Enhancement of supersymmetry near black hole horizons}, Phys. Rev.
{\bf D55} (1997) 3647;hep-th/9610155}
\REF\sabratwo{ A. Chamseddine and W.A. Sabra, {\sl Calabi-Yau black holes
and enhancement of supersymmetry in five dimensions},  Phys. Lett. 
{\bf B460} (1999) 63; hep-th/9903046.}
\REF\coles { R.A.Coles and G.Papadopoulos,{\sl The
Geometry of one-dimensional supersymmetric
nonlinear sigma models}, Class. Quantum Grav.
{\bf 7} (1990) 427.}
\REF\renata {G.W.Gibbons and R.Kallosh,{\sl
Topology,entropy and the Witten index of
dilaton black holes}, Phys. Rev. {\bf D51}
(1995) 2839.}
\REF\gary{G.W. Gibbons, G. Papadopoulos and K. Stelle, {\sl HKT and OKT
Geometries on Soliton Black Hole Moduli Spaces}, Nucl. Phys. {B508} (1997) 623,
hep-th/9706207.}
\REF\twist{P.S. Howe and G. Papadopoulos, {\sl Twistor Spaces
for HKT Manifolds}, Phys. Lett. {\bf B379} (1996)80, hep-th/9602108.}
\REF\poon{G. Grantcharov and Y.S. Poon, {\sl Geometry of hyper-K\"ahler
Connections with Torsion}, math.dg/9908015.}
\REF\micha{J. Michelson and A. Strominger, {\sl Superconformal 
Multi-Black Hole
Quantum Mechanics}, HUTP-99/A047, hep-th/9908044.}
\REF\fere {R.C.Ferrell \& D.M.Eardley,{\sl
Slow motion scattering and coalescence of
maximally charged black holes},Phys. Rev.
Lett. {\bf 59} (1987) 1617.}
\REF\shiraishi{K. Shiraishi, {\sl Moduli Space Metric for Maximally-Charged
Dilaton B;ack Holes}, Nucl. Phys. {\bf B402} (1993) 399.}
\REF\gibr {G.W.Gibbons \& P.J.Ruback,{\sl
The motion of extreme Reissner-Nordstrom black
holes in the low velocity limit},Phys. Rev.
Lett. {\bf 57} (1986) 1492.}
\REF\michb{J. Michelson and A. Strominger, {\sl The 
geometry of (Super)conformal
Quantum Mechanics}, HUTP-99/A045, hep-th/9907191.}
\REF\tesc{G. Papadopoulos and A. Teschendorff, {\sl Multi-angle
Five-Brane Intersections}, Phys. Lett. {\bf B443} (1998) 
159, hep-th/9806191;
{\sl Grassmannians, Calibrations and
 Five-Brane Intersections}, hep-th/9811034.}
\REF\cremmer{E. Cremmer, {\sl Supergravities in Five Dimensions}, in
\lq\lq Superspace and  Supergravity", eds. S.W. Hawking and M. Ro\v cek
CUP 1981, page 267.}
\REF\stu{I. Antoniadis, S. Ferrara and T.R. Taylor, {\sl N=2
Heterotic Superstrings and its Dual Theory in Five Dimensions},
 Nucl. Phys. {\bf B460} (1996)
489; hep-th/9511108.}
\REF\sabrac{I. Gauda, S. Mahapatra, T. Mohaupt and W. A. Sabra,
 {\sl Balck Holes and
Flop Transitions in M-theory on Calabi-Yau Threefolds}, Class. Quantum
Grav. {\bf 16} (1999) 419; hep-th/9807014.}
\REF\kalloshb{R. Kallosh, A. Linde and M. Shmakova, {\sl Supersymmetric
Multiple Basin Attractors}, hep-th/9910021.}


\date{October 1999}
\titlepage
\title{The Dynamics of Very Special Black Holes }
\author{J. Gutowski }
\address{DAMTP,  University of Cambridge,\break
 Silver Street, Cambridge CB3 9EW}
 \andauthor{G. Papadopoulos}
 \address{Department of Mathematics, King's College London,\break
 Strand, London WC2R 2LS}
 \abstract{We show that the moduli space of supersymmetric   black holes that
 arise in the five-dimensional $N=2$ supergravity theory 
 with  any number
 of vector multiplets
 is a weak  HKT manifold. The moduli metric
  is expressed in terms of a HKT potential which is 
 determined by the associated very special geometry 
 of the supergravity theory. As an example, we give explicitly the black
hole moduli metric for the STU model.}

\endpage

\pagenumber=1




\chapter{Introduction}

In the last few years, there has been much interest in   
supersymmetric  black holes in five dimensions.
This is because of the 
Strominger and Vafa [\strom] 
microscopic derivation of the Bekenstein-Hawking entropy
for a class of such black holes. 
The low energy description of M-theory compactifications
on  six-dimensional Calabi-Yau manifolds [\ferrara, \gpapas] and heterotic string 
theory compactifications
on $K_3\times S^1$ [\grwest]  is given by  $N=2$ supersymmetric
five-dimensional supergravity
theories (eight supercharges) coupled to 
vector and scalar multiplets [\town]. 
A novel property of the $N=2$ supergravity 
theories, as shown by B. de Wit and 
 A. van Proeyen in [\wit], 
 is that the couplings of the vector multiplets can be determined
 in terms of geometric data on very special manifolds.
These theories admit  electrically charged
supersymmetric black hole solutions which preserve $1/2$ 
of supersymmetry. Sabra and Chamseddine in [\sabra] have shown 
that these black holes can be  described  in terms
of the underlying   very special 
geometry.

Viewing the supersymmetric black holes as the BPS soliton 
solutions of  supergravity theories,  their low energy dynamics
can be approximated by  geodesic motion in a
moduli space. Some  black hole properties like low energy scattering
and the presence of bound states can be investigated by studying
the geometric structure of the moduli spaces.
In turn these may have applications in the understanding
of M-theory and its compactifications and perhaps provide
another way of deriving the Bekenstein-Hawking entropy formula. 
Recently it has been realized that the black hole moduli
spaces exhibit geometric structures that are related to those that
appear on the target spaces of one-dimensional supersymmetric
sigma models [\coles]. This is because the low energy effective
theory of black holes,  which is an one-dimensional sigma model, has as many
supersymmetries as those preserved by the associated solutions [\renata, \gary].
  An example of such geometry is
that  of hyper-K\"ahler with torsion (HKT) [\twist] which arises
in a class of one-dimensional sigma models with four
supersymmetries based on the $N=4B$ multiplet; for  applications in
mathematics see Grantcharov and Poon [\poon]. 
In particular, it has been shown in [\gary] that the moduli
space of five-dimensional black holes that preserve $1/4$
of the maximal supersymmetry is a strong HKT manifold. 
Michelson  and Strominger in [\micha] extended this
to five-dimensional black holes which preserve $1/8$ of
supersymmetry  and  are electrically charged
with respect to the gauge vector potential of the 
supergravity multiplet. 
In particular, they established that the  
moduli space of these black holes
admits a weak HKT structure.

In this paper, we investigate the moduli space of electically
charged black holes of five-dimensional $N=2$ supergravity  with
any number of vector multiplets. We show that the moduli space
is a weak HKT manifold. This is in agreement with
the counting of the number of unbroken supersymmetries of
the black hole solutions and the expected $N=4B$ multiplet structure
of the effective theory.
The HKT metric can be expressed in terms 
of a HKT potential
$$
\mu=\int d^4x\, e^{6U}
\eqn\inone
$$
where $U$ is determined by the very special
geometry of the supergravity theory. In this way, we establish a
relation between the very special geometry that arises in
 $N=2$ supergravity theory
and the weak HKT geometry that arises in one-dimensional 
supersymmetric sigma models.
 We  give as an
explicit example
the moduli space metric of the black holes that arise in
STU model associated with
the $K_3\times S^1$ compactification of the heterotic string.

This paper is organized as follows: In section two,
we describe the  supersymmetric electrically charged
black hole solutions of $N=2$ five-dimensional supergravity theory
and establish our notation. In section three, we give
the black hole moduli metric and show that it is weak HKT.
In section four, we present an example associated with
the STU model and in section five we give our conclusions.

\chapter{The Black Holes of Five-Dimensional Supergravity }

In this section we shall review some facts about very special
geometries and their applications to five-dimensional supergravity.
The bosonic part action of
five-dimensional
$N=2$ supergravity  with $k$ vector multiplets is associated to  a
 hypersurface $N$ of 
$\bR^k$ defined by the equation 
$$ 
V(X) \equiv {1 \over 6} C_{IJK}X^I 
X^J X^K=1 
\eqn\bone
$$ 
where $\{X^I; I=1, \dots, k\}$ are standard  coordinates 
on $\bR^k$ and $C_{IJK}$ are constants. In the case of a
model arising from a Calabi-Yau compactification of M-theory, the constants 
$C_{IJK}$ are the topological intersection numbers of the compact manifold.
Next we set
$$
\eqalign{
Q_{IJ}&\equiv-{1 \over 2}{\partial \over \partial X^I}{\partial \over \partial
X^J} \log V \mid_{V=1}
={9 \over 2}X_I X_J - {1 \over 2}C_{IJK}X^K
\cr
h_{ab} &= Q_{IJ} {\partial X^I \over \partial \f^a}{\partial X^J \over \partial
\f^b} \mid_{V=1}\ ,}
\eqn\btwo
$$
where $\{\phi^a; i=1,\dots, k-1\}$ are local
 coordinates of $N$, $h$ is interpreted as  a metric
on $N$ and
$$
X_I ={1 \over 6}C_{IJK}X^J X^K
\eqn\bthree
$$
are the dual coordinates to $X^I$. Note that the hypersurface equation $V=1$
can also be rewritten as  $X^I X_I =1$.
Then, the bosonic part of the associated supergravity  action [\wit]
with vector potentials $A^I$ and scalars $\phi^a$ is  
$$
\eqalign{
S =& \int d^5 x \sqrt{-g} \big[ R +{1 \over 2}Q_{IJ}{F^I}_{\m
\n}{F^J}^{\m \n}+h_{ab}\pd{\m} \f^a \pu{\m} \f^b\big] 
\cr &-{1 \over
24}e^{\m \n \p \s \t}C_{IJK}{F^I}_{\m \n}{F^J}_{\p \s}{A^K}_\t}
\eqn\acti
$$
where $F^I = d A^I$, $I,J,K=1,\dots, k$ are the 2-form Maxwell field strengths,
$\mu, \nu, \rho, \sigma=0, \dots, 4$,
and $g$ is the metric of the five-dimensional spacetime;
we have used the same symbol $\phi^a$ to denote both
the coordinates of $N$ and the various scalar fields of the theory. 
 
The field equations of the above Lagrangian obtained from varying
the scalars $\f^a$, the spacetime metric $g$, and the vector
potentials $A^I$  are
$$
\eqalign{
\sqrt{-g} \partial_a Q_{IJ} \big[ {1 \over 2}{F^I}_{\m \n}{F^J}^{\m
\n}+\pd{\m}X^I \pu{\m}X^J \big] - 2 \pd{\m} \big( \sqrt{-g}Q_{IJ}
\pu{\m}X^I \big) \partial_a X^J =0\ ,}
\eqn\feqa
$$
$$
\eqalign{
\sqrt{-g} \big( G_{\m \n}+Q_{IJ}{F^I}_{\m \p}{{F^J}_{\n}}^\p
+Q_{IJ} \pd{\m}X^I \pd{\n}X^J \big)
\cr
-{1 \over 2}\sqrt{-g}g_{\m \n} \big[ {1 \over 2}Q_{IJ} {F^I}_{\p
\s}{F^J}^{\p \s} +Q_{IJ}\pd{\p}X^I \pu{\p}X^J \big]=0\ ,}
\eqn\feqb
$$
and
$$
\eqalign{
-2 \pd{\m} \big[ \sqrt{-g} Q_{IJ} {F^J}^{\m \n} \big] -{1 \over
8}e^{\n \p \s \m \t}C_{IJK} {F^J}_{\p \s}{F^K}_{\m \t} =0\ ,}
\eqn\feqc
$$
respectively.
The electrically charged black hole solutions [\sabra]  that preserve $1/2$ 
of supersymmetry of $N=2$ supergravity action \acti\  are 
$$
\eqalign{
ds^2 &= -e^{-4U} dt^2 +e^{2U} d {\bf{x}}^2
\cr
{A^I}_0 &= e^{-2U} X^I
\cr
e^{2U}X_I& ={1 \over 3}  H_I\ ,}
\eqn\sol
$$
where
$$
H_I = h_I + \sum_{A=1}^{N_I} {{\l_{IA}} \over
|{\bf{x}}-{\bf{y}}_{IA}|^2}
\eqn\ham
$$
is a harmonic function on $\bR^4$ with $N_I$ centres.  Viewing $e^U$
as an additional scalar, the last equation in \sol\
gives the  $k$ independent scalars $\{e^U, \phi^a\}$ in terms of the $k$ harmonic
functions $\{H_I\}$. In what follows,  we shall assume that the black hole
solutions exist, i.e. that the scalars and the components of
the metric can be expressed in terms of the harmonic functions. However,
this depends on the existence of solutions of the
 stabilization equations (see [\kalloshc]).

A special class of solutions are those for which the positions of
the different harmonic functions are the same\foot{We thank A. Strominger
for pointing this out to us.}, i.e. ${\bf y}_{IA}={\bf y}_{JA}$
for $I\not= J$. The associated black hole solutions are of 
interest since they exhibit regular horizons [\cfgk, \sabratwo]. The 
moduli metric of these black holes can be easily found as a special case
of that of the more general solution \sol\ above.

The source term associated to the solution
\sol\ is
$$
S_{source} =2 V_3 \int d^4x \sum_{I,A} d \t_{IA} \d
({\bf{x}}-{\bf{y}}_{IA})\big( X^I \l_{IA} - {A^I}_\m \l_{IA} {d
{y_{IA}}^{\m} \over d \t_{IA}} \big)\ ,
\eqn\sour
$$
where $V_3$ is the volume of the unit three sphere,
$$
d\tau_{IA}=\sqrt{-g_{\mu\nu} {dy^\mu_{IA}\over dt} {dy^\nu_{IA}\over dt}}\, dt
\eqn\bsour
$$
and we have set $y^0_{IA}=t$. The addition 
of \sour\ is due to the presence of non-vanishing delta function sources
as $|{\bf x}-{\bf y}_{IA}|\rightarrow 0$.

\chapter{The moduli metric}

The moduli space of the black holes \sol\ reviewed
in the previous section is expected to be a weak HKT manifold.
This is because that although these solutions \sol\ 
include those that have been used in [\micha] as a special
case, in both cases the killing spinor equations 
impose to the same conditions
on the killing spinors. Consequently, in both 
cases the sigma models 
that prescribe
the low energy dynamics have the same type of multiplet.
The relevant multiplet in this case is the $N=4B$ which is associated
with the weak HKT geometry [\coles, \gary].  
In many moduli problems, much of the geometric structure on the moduli
space is induced from the geometric structure of the underlying space(time).
For the black holes \sol, the spatial transverse space is $\bR^4$
and so it admits a (constant) hypercomplex structure. The hypercomplex
structure of the HKT moduli space is induced from that of $\bR^4$.

To compute the metric on the black hole moduli space,
we follow [\fere, \shiraishi, \gibr] and  allow the positions ${\bf y}_{IA}$
to depend on time, i.e.
$$
{\bf y}_{IA}\rightarrow {\bf y}_{IA}(t)\ .
\eqn\cone
$$ 
In addition, we perturb the metric and
the gauge potentials as
$$
\eqalign{
ds^2&\rightarrow ds^2+ 2 e^{-4U} p_m dt dx^m
\cr
A^I_0dt&\rightarrow A^I_0 dt+ ({D^I}_m-e^{-2U}X^I p_m) dx^m}
\eqn\ana
$$
where we take $p_m$ and ${D^I}_m$ to be first order in the velocities
and we have set $x^\mu=(t, x^m)$, $m,n=1,\dots, 4$. The scalar
fields are not perturbed linear in the velocities. The
$p_m$ and ${D^I}_m$ will be determined by solving the supergravity
field equations.

Next we solve the supergravity equations taking into account
the source terms to first
order in the velocities. The relevant equations are those
of the gauge vector potential and those of the metric. The
field equation of the scalars vanishes to the first order in
the velocities. For convenience, we define
$$
\eqalign{
f_{mn}&=\partial_m p_n-\partial_n p_m
\cr
f^I_{mn}&=\partial_m D_n^I-\partial_n D_m^I\ .}
\eqn\ctwo
$$
Substituting the ansatz \ana\ into the
field equations and collecting the terms linear in the velocities,
we find
$$
\eqalign{
\pd{0}\pu{n} H_I +3 \pd{m} \big( e^{-4U}X_I f^{mn}& \big)
\cr
 - 2 \pd{m} \big( e^{-2U}Q_{IJ} f^{Jmn} \big)
-{1 \over 2}&\e^{nr \ell m}\pd{r} \big(C_{IJK}e^{-2U}X^K \big)
 f^{J}{}_{\ell m})
\cr
+{3 \over 2} \e^{nr \ell m}
\pd{r} \big(e^{-4U} X_I &\big)f_{\ell m}=2V_3 \sum_{A} \l_{IA}
\d({\bf{x}}-{\bf{y}}_{IA}) {v_{IA}}^n}
\eqn\pqn
$$
$$
\eqalign{
e^{2U} \big[ {1 \over 2}e^{-2U}X^I \pd{0}\pd{n}H_I +{1 \over
2}e^{-6U} &\pu{r}(H_J) f^{J}{}_{nr}
\cr
+{1 \over 2}e^{-6U} \pu{m}\big( f_{mn}
\big) \big]&=V_3 \sum_{I,A}X^I \l_{IA} \d
({\bf{x}}-{\bf{y}}_{IA}) {v_{IA \ n}}\ ,}
\eqn\peqn
$$
where indices are raised and lowered with respect to the
Euclidean metric on $\bR^4$.
To derive the above equations, we remark that from the field equations
of the metric and those of the gauge field only the $0n$ and the $n$ 
components contribute, respectively. We have also used the
identities
$$
\pd{0}\pd{n}(e^{2U})={1 \over 3}X^I \pd{0}\pd{n}H_I
+e^{2U}\pd{n}X_I \pd{0}X^I\ .
\eqn\mone
$$
Next we contract the field equation  \pqn\ with ${1\over2} X^I$ and subtract it from \peqn. This gives
$$
\eqalign{
\pd{m} \big[ e^{-6U}\big(f^{mn} -{1 \over 2}
\e^{mnr \ell}f_{r\ell} \big)
\cr
-{3 \over 2}X_I e^{-4U}\big( f^{Imn}-{1 \over
2}\e^{mnr \ell}f^I_{r\ell}\big) \big]=0\ .}
\eqn\mtwo
$$
Solving this equation for $p$, we find
$$
\eqalign{
p^n = -{1 \over {2V_3}} \int d^4 z {1 \over |{\bf{x}}-{\bf{z}}|^2}
\pd{r} \big[{3 \over 2}e^{2U}X_I
\big(f^{Irn}-{1\over2}\e^{rn \ell s}f^I_{\ell s}\big)
\big]\ .}
\eqn\sola
$$
Substituting the expression for $p$ back into  \pqn, we find
$$
\eqalign{
D^{Jn}=-{1 \over V_3} \int d^4 z {1 \over |{\bf{x}}-{\bf{z}}|^2}
\pd{r} \big[ {{B}}^{JI} 
\big(\pu{r}{K_I}^n - \pu{n}{K_I}^r+\e^{rn \ell
s}\pd{\ell}K_{Is} \big) \big]\ ,}
\eqn\solb
$$
where ${{B}}^{JI}$ is the inverse of the matrix
$$
{{B}}_{IJ}=e^{-2U} (2C_{IJK}X^K -9X_I X_J)
\eqn\mthree
$$
and
$$
{K_I}^n = - \sum_A {\l_{IA} {v_{IA}}^n \over
|{\bf{x}}-{\bf{y}}_{IA}|^2}\ .
\eqn\mfour
$$

It remains to calculate the moduli metric. For this, we must substitute 
the solutions \sola\ and \solb\ into the action including the source
term and collect the terms $S^{(2)}$ and $S^{(2)}_{\rm source}$ which
are quadratic in the
velocities up to surface terms. The second order term in the 
source free action is
$$
\eqalign{
S^{(2)}=&\int d^5x\, \big [3 e^{2U} (\pd{0} e^{2U})^2 
+{1 \over 2}e^{-6U}f_{mn}f^{mn}
+{3 \over 2}e^{6U} \pd{0}X_I \pd{0}X^I
- (\pd{0}{D^I}_n)\pu{n}H_I
\cr
+&{1 \over 2}e^{-2U}Q_{IJ}f^I_{mn}f^{Jmn}
-{3 \over 2}e^{-4U}X_I f^I_{mn}f^{mn}
-{1 \over 8}e^{-2U}C_{IJK}X^K \e^{mnr \ell}f^I_{mn}f^J_{r\ell}
\cr
+&{3 \over 4}e^{-2U}X_I \e^{m n r \ell}f^I_{mn}f_{r\ell}
-{1 \over 4}e^{-6U}\e^{mnr \ell}f_{mn}f_{r\ell}\big ]}
\eqn\mfive
$$
and upon  substituting the solution gives
$$
S^{(2)}=
-\int d^5x\, {1 \over 2}e^{4U}X^I \pd{0}\pd{0}H_I +{1 \over 8}{
{B}}_{IJ} \big[ f^I_{mn}f^{Jmn}
+{1 \over 2} \e^{mnr \ell}
f^I_{mn}f^J_{r\ell}\big] \ .
\eqn\qlaga
$$
The second order contribution from the source terms is
$$
S_{\rm source}^{(2)}=- V_3 \int d^5x \sum \d ({\bf{x}}-{\bf{y}}_{IA}) \big[X^I
\l_{IA} |{\bf{v}}_{IA}|^2 e^{4U} +2 {D^I}_m \l_{IA} {v_{IA}}^m \big]\ .
\eqn\qlagb
$$
Adding \qlaga\ and \qlagb\ and  using
$$
\eqalign{
- 2\int d^5x \, V_3 \sum {D^I}_m \l_{IA} &{v_{IA}}^m \d
({\bf{x}}-{\bf{y}}_{IA})
\cr
=& {1 \over 4}\int d^5x\, {
{B}}_{IJ} \big[ f^I_{mn}f^{Jmn}
+{1 \over 2} \e^{mnr \ell}
f^I_{mn}f^J_{r\ell}\big]}
\eqn\msix
$$
and
$$
\eqalign{
{{B}}^{IJ}X_J = {1 \over 3}e^{2U} X^I
\cr
{{B}}^{IJ} \pd{\m} X_J = {1 \over 6}e^{2U} \pd{\m} X^I
\cr
{\pd{\m}{{B}}^{IJ}}X_J = {1 \over 3}{\pd{\m}e^{2U}}X^I
+{1 \over 2}e^{2U} \pd{\m}X^I\ ,}
\eqn\mseven
$$
we find that
$$
\eqalign{ S^{(2)}&+S_{\rm source}^{(2)}=
\cr 
\int d^5x &\big[-V_3 \sum X^I \l_{IA} |{\bf{v}}_{IA}|^2 e^{4U} \d
({\bf{x}}-{\bf{y}}_{IA}) - K_{Jn} \pd{r} \big( {
{B}}^{IJ} \pu{r}{K_I}^n \big)
\cr
+&{ {B}}^{IJ}\big( \pd{0}H_I \pd{0}H_J -
\pd{r}K_{Jn}\pu{n}{K_I}^r \big) +{ {B}}^{IJ} \e^{r n \ell
s} \pd{r}K_{Jn}\pd{\ell}K_{Is}\big]\ .}
\eqn\main
$$
>From this we can easily read the moduli metric which turns out to
be weak HKT. To see this, let $(I_{s})$ be the triplet of constant  complex
structures on $\bR^4$ satisfying the algebra of the imaginary
unit quaternions which are associated to self-dual 2-forms on
$\bR^4$. Then
$$
g_{m IA \ n J B}= \pd{m I A}\pd{n J B}\, \mu +\sum_{s=1}^3
{(I_s)}^\ell{}_{m} (I_s)^q{}_n \pd{\ell I A} \pd{q J B}\,\mu
\eqn\modmet
$$
where 
$$
\mu= \int d^4 x\, e^{6U}
\eqn\mmone
$$
is the HKT potential  and $e^U$ 
given in \sol; for a discussion about HKT potentials see [\michb, \poon]. 
To show this, we have used
$$
\sum_{s=1}^3
{(I_s)}^\ell{}_{m} (I_s)^q{}_n = \d_{m n} \d^{\ell
q}-{\d_m}^q {\d_n}^\ell -{\e_{m n}}^{\ell q}\ ,
\eqn\mmtwo
$$
$$
\pd{m I A} (e^{2U}) = {1 \over 3}X^I \pd{m I A} H_I\ ,
\eqn\mmthree
$$
(no sum over $I$) and
$$
\pd{m J A} X^I =-2 e^{-4U} { {B}}^{IJ} \pd{m}
\big({\l_{J A} \over |{\bf{x}}-{\bf{y}}_{J A}|^2} \big) +{2
\over 3}e^{-2U} X^I X^J \pd{m}
\big({\l_{J A} \over |{\bf{x}}-{\bf{y}}_{J A}|^2} \big)\ .
\eqn\mmfour
$$
(no sum over $I$ and $J$). For a generic choice of very special geometry
and a generic choice of a black hole solution, the torsion of the
HKT geometry is not a closed form. So the moduli space of the
$N=2$ supergravity black holes is a weak HKT manifold with metric
\modmet\ and hypercomplex structure $\{I_s\}$. We remark that the 
moduli metric  \modmet\ has a term symmetric  and a term anti-symmetric 
in the spatial spacetime indices. The anti-symmetric piece
can be written in a basis of anti-self-dual two-forms in $\bR^4$.
Such HKT geometries have been considered in the past  [\tesc].

The moduli metric of the black holes associated with harmonic functions
which have the same positions is also given by \mmone. However in the
expression for the HKT potential the corresponding harmonic functions are used.
The moduli metric is again given by \modmet\ but the derivatives are taken
with respect to the independent positions ${\bf y}_A$ of the harmonic functions.

\chapter{Examples}

One possibility is to consider black holes that  are coupled 
 to a single one-form gauge potential.
For this we choose 
$$
C_{111}=1\ .
\eqn\onon
$$
The corresponding action \acti\ is that of a pure five-dimensional
supergravity multiplet which has bosonic fields   a 
graviton and an one-form gauge potential [\cremmer].
 We remark that our metric in this case is in agreement
with that of [\micha].

Alternatively, we can consider the moduli space of black holes
that are coupled to different one-form gauge potentials. 
To give an example,  we shall describe in
detail the metric on the moduli space of black holes of the $STU$ model.
This model arises in the context of compactifications of the
heterotic string on $K_3\times S^1$ and the associated very special
geometry has been presented in [\stu]. In this case, 
we take $I,J, K =1,2,3$ and the non-vanishing 
component of $C$ is
$$
C_{123}=1\ .
\eqn\onbon
$$
Then,  $X^I$ and  $e^{2U}$ are expressed in terms of the harmonic
functions as
$$
\eqalign{
e^{2U}&=(H_1 H_2 H_3)^{1 \over 3}
\cr
X^1&= \big( {H_2 H_3 \over {H_1}^2} \big)^{1 \over 3}
\cr
X^2&= \big( {H_1 H_3 \over {H_2}^2} \big)^{1 \over 3}
\cr
X^3&= \big( {H_1 H_2 \over {H_3}^2} \big)^{1 \over 3}}
\eqn\nana
$$
Similarly, the non-vanishing components of ${ {B}}^{IJ}$ are
$$
\eqalign{
{ {B}}^{12} &= {1 \over 2}H_3
\cr
{ {B}}^{13} &= {1 \over 2}H_2
\cr
{ {B}}^{23} &= {1 \over 2}H_1\ .}
\eqn\nbna
$$
Substituting these into the expression for the moduli metric, we find
$$
\eqalign{
ds^2=
-&V_3 \big[h_2 h_3 \sum_A \l_{1A} |d{\bf{y}}_{1A}|^2+h_1 h_3 \sum_A
\l_{2A} |d{\bf{y}}_{2A}|^2+h_1 h_2 \sum_A \l_{3A} |d{\bf{y}}_{3A}|^2
\big]
\cr
-V_3 h_2 &\sum_{A,B} { \l_{1A} \l_{3B} \over
|{\bf{y}}_{1A}-{\bf{y}}_{3B}|^2} |d{\bf{y}}_{1A} - d{\bf{y}}_{3B}|^2
-V_3 h_1 \sum_{A,B} { \l_{2A} \l_{3B} \over 
|{\bf{y}}_{2A}-{\bf{y}}_{3B}|^2} |d{\bf{y}}_{2A} - d{\bf{y}}_{3B}|^2
\cr 
-&V_3 h_3 \sum_{A,B} { \l_{1A} \l_{2B} \over
|{\bf{y}}_{1A}-{\bf{y}}_{2B}|^2} |d{\bf{y}}_{1A} - d{\bf{y}}_{2B}|^2
\cr
-&{1 \over 2}V_3 \sum_{A,B,C} \l_{1A} \l_{2B}\l_{3C}
|d{\bf{y}}_{1A}-d{\bf{y}}_{2B}|^2 \big[ {1 \over
|{\bf{y}}_{1A}-{\bf{y}}_{3C}|^2|{\bf{y}}_{1A}-{\bf{y}}_{2B}|^2}
\cr
&\qquad\qquad+{1 \over
|{\bf{y}}_{2B}-{\bf{y}}_{3C}|^2|{\bf{y}}_{2B}-{\bf{y}}_{1A}|^2}
-{1 \over
|{\bf{y}}_{1A}-{\bf{y}}_{3C}|^2|{\bf{y}}_{2B}-{\bf{y}}_{3C}|^2}\big]
\cr
-&{1 \over 2}V_3 \sum_{A,B,C} \l_{1C} \l_{2A}\l_{3B}
|d{\bf{y}}_{2A}-d{\bf{y}}_{3B}|^2 \big[ {1 \over
|{\bf{y}}_{2A}-{\bf{y}}_{1C}|^2|{\bf{y}}_{2A}-{\bf{y}}_{3B}|^2}
\cr
&\qquad\qquad+{1 \over
|{\bf{y}}_{3B}-{\bf{y}}_{1C}|^2|{\bf{y}}_{3B}-{\bf{y}}_{2A}|^2}
-{1 \over
|{\bf{y}}_{2A}-{\bf{y}}_{1C}|^2|{\bf{y}}_{3B}-{\bf{y}}_{1C}|^2}\big]
\cr
-&{1 \over 2}V_3 \sum_{A,B,C} \l_{1A} \l_{2C}\l_{3B}
|d{\bf{y}}_{1A}-d{\bf{y}}_{3B}|^2 \big[ {1 \over
|{\bf{y}}_{1A}-{\bf{y}}_{2C}|^2|{\bf{y}}_{1A}-{\bf{y}}_{3B}|^2}
\cr &
\qquad\qquad+{1 \over
|{\bf{y}}_{3B}-{\bf{y}}_{1A}|^2|{\bf{y}}_{3B}-{\bf{y}}_{2C}|^2}
-{1 \over
|{\bf{y}}_{1A}-{\bf{y}}_{2C}|^2|{\bf{y}}_{3B}-{\bf{y}}_{2C}|^2}\big]
\cr
+&2 \int d^4 x \sum_{A,B,C} {\l_{1C} \l_{2A} \l_{3B} \over
|{\bf{x}}-{\bf{y}}_{1C}|^2} \big({dy_{2A}}^{[m} {dy_{3B}}^{n]}\big)^-
\pd{m} \big({1 \over |{\bf{x}}-{\bf{y}}_{2A}|^2}\big) \pd{n}
\big({1 \over |{\bf{x}}-{\bf{y}}_{3B}|^2} \big)  
\cr
+ &2\int d^4 x \sum_{A,B,C} {\l_{2C} \l_{3A} \l_{1B} \over
|{\bf{x}}-{\bf{y}}_{2C}|^2} \big({dy_{3A}}^{[m} {dy_{1B}}^{n]}\big)^- 
\pd{m} \big({1 \over |{\bf{x}}-{\bf{y}}_{3A}|^2}\big) \pd{n}
\big({1 \over |{\bf{x}}-{\bf{y}}_{1B}|^2}\big)  
\cr
+& 2\int d^4 x \sum_{A,B,C} {\l_{3C} \l_{1A} \l_{2B} \over
|{\bf{x}}-{\bf{y}}_{3C}|^2} \big({dy_{1A}}^{[m} {dy_{2B}}^{n]}\big)^-  
\pd{m} \big({1 \over |{\bf{x}}-{\bf{y}}_{1A}|^2}\big) \pd{n}
\big({1 \over |{\bf{x}}-{\bf{y}}_{2B}|^2} \big) }
\eqn\stumod
$$
where
$$
\big({dy_{2A}}^{[m} {dy_{3B}}^{n]}\big)^-=
{dy_{2A}}^{[m} {dy_{3B}}^{n]}-{1\over2}
\epsilon^{mn}{}_{rs} {dy_{2A}}^{[r} {dy_{3B}}^{s]}
\eqn\lasrr
$$
and similarly for the rest. Observe that the moduli metric
has the general structure of HKT metrics considered in [\tesc].

The moduli metric indicates that there are up to three body interactions.
One possible explanation for this is that these black holes are in the
same universality class as the black holes that are made  from {\sl three}
intersecting branes in the case of toroidal 
compactifications of M-theory to five dimensions. In both cases
the black holes preserve the same fraction of maximal supersymmetry.
So the three body interactions reflect the fact that the black holes
are made from three different objects.

Now consider the case of three black holes each coupled to a different
one-form gauge potential.  One might have
expected that the moduli metric simplifies in this case in analogy
with a similar situation in the context of BPS solitons but this
does not seem to be the case here. This might be due to the
fact that the two scalars involved in the model have non-trivial
 interactions in the action.
 
Finally, the moduli metric of STU  black holes for which the associated three
 harmonic
functions  have the same positions, i.e. ${\bf y}_{1A}={\bf y}_{2A}=
{\bf y}_{3A}={\bf y}_{A}$, is
$$
\eqalign{
ds^2&=
-V_3 \sum_A \big[h_2 h_3  \l_{1A} +h_1 h_3
\l_{2A} +h_1 h_2  \l_{3A}
\big] |d{\bf{y}}_{A}|^2
\cr
-&V_3  \sum_{A \neq B} \big[ h_2 \l_{1A} \l_{3B} \
+ h_1  \l_{2A} \l_{3B}
+ h_3  \l_{1A} \l_{2B} \big] {|d{\bf{y}}_{A}-d{\bf{y}}_{B}|^2 \over
|{\bf{y}}_{A} -
{\bf{y}}_{B}|^2}
\cr
-&{1 \over 2}V_3 \sum_{A \neq B,C} \big[ \l_{1A} \l_{2B}\l_{3C}+\l_{1C}
\l_{2A}\l_{3B}
+ \l_{1A} \l_{2C}\l_{3B} \big]
|d{\bf{y}}_{A}-d{\bf{y}}_{B}|^2
\cr
&\big[ {1 \over
|{\bf{y}}_{A}-{\bf{y}}_{C}|^2|{\bf{y}}_{A}-{\bf{y}}_{B}|^2}
+{1 \over
|{\bf{y}}_{B}-{\bf{y}}_{C}|^2|{\bf{y}}_{B}-{\bf{y}}_{A}|^2}
-{1 \over
|{\bf{y}}_{A}-{\bf{y}}_{C}|^2|{\bf{y}}_{B}-{\bf{y}}_{C}|^2}\big]
\cr
+&2 \sum_{A \neq B \neq C} \int d^4 x \big[\l_{1C} \l_{2A} \l_{3B}
+\l_{2C} \l_{3A} \l_{1B}
+\l_{3C} \l_{1A} \l_{2B}\big]
\cr
&{\big({dy_{A}}^{[m} {dy_{B}}^{n]}\big)^- \over |{\bf{x}}-{\bf{y}}_{C}|^2}\,\,
\pd{m} \big({1 \over |{\bf{x}}-{\bf{y}}_{A}|^2}\big) \pd{n}
\big({1 \over |{\bf{x}}-{\bf{y}}_{B}|^2} \big)\ . }
\eqn\stumodb
$$

More recently, new examples of five-dimensional black holes
have been found [\sabrac] following the existence of new  
solutions of the stabilization
equations in [\kalloshc, \kalloshb]. The black hole moduli metric can be
computed in this case as well. It would be of interest to find
how the moduli metric changes in  the various transitions. 

\chapter{Concluding Remarks}
We have computed the moduli metric of five-dimensional black
holes of $N=2$ supergravity coupled to any number of vector
multiplets. We have found that the moduli space is a weak
HKT manifold.
One can investigate the near \lq horizon limit' of our moduli metric
as in [\micha]. In particular, we examined
the near horizon limit of the STU model moduli metric \stumod\  that
we have presented in the previous section. We found though
that the moduli metric appears to be singular in this limit. 
This may be due to fact that the associated black holes
do not have a regular horizon. However, it is expected that
some examples of moduli metrics will exhibit a well defined
near horizon limit especially those for which all the harmonic
functions have the same positions.

To investigate black hole scattering and  black hole bound states
a new way to construct these metrics may be required. Although
the moduli space of $n$ black holes can be identified with the
configuration space of $n$ particles, the computation of the metric
is local. A different construction may provide the tools to find
closed geodesics in the black hole moduli spaces and so
investigate the presence of black hole bound states.
\vskip 0.5cm
{\bf Acknowledgments:} We thank S. Ferrara, R. Kallosh and A. Strominger for
 helpful suggestions
and comments. J.G. thanks EPSRC for a studentship. G.P. is
supported by a University Research Fellowship from the Royal Society.
\refout

\end